# Effects of randomization of characteristic times on spiral wave generation in a simple cellular automaton model of excitable media

Vincent Vangelista, Karl Amjad-Ali, Minhyeok Kwon, and Paulo H. Acioli[1]

*Department of Physics*

*Northeastern Illinois University*

*5500 St. Louis Ave, Chicago, IL 60625*

Abstract

.

Spiral waves are self-repeating waves that can form in excitable media, propagating outward from their center in a spiral pattern. Spiral waves have been observed in different natural phenomena and have been linked to medical conditions such as epilepsy and atrial fibrillation. We used a simple cellular automaton model to study propagation in excitable media, with a particular focus in understanding spiral wave behavior. The main ingredients of this cellular automaton model are an excitation condition and characteristic excitation and refractory periods. The literature shows that fixed excitation and refractory periods together with specific initial conditions generate stationary and stable spiral waves. In the present work we allowed the activation and refractory periods to fluctuate uniformly over a range of values. Under these conditions formed spiral waves might drift, the wave front might break, and in some extreme cases it might lead to a complete breakdown of the spiral pattern.

## I. Introduction

Spiral waves have been observed in natural systems for many decades. These observations include spiral patterns in the Belousov-Zhabotinsky reaction[1], the impact of spiral waves on heart fribilation,[2] and spiral growth pattern in crytals.[3] The motivation on the present work was the recently published article of Welsh *et al.*,[4] in which the authors gathered 500-600 participants to simulate spiral wave patterns. In their experiment they set up a grid of participants and provided them with rules that mimic the transmission of electric signals in a network of neurons. The stimulus was simulated by the act of a participant raising their hands for a short period time if they had two of their immediate neighbors with a raised hand, and after, to keep their hands down for a short refractory period. They were able to generate spiral waves by starting some of the participants in a line on an initial refractory state (hands down for a period of time) next to a line of participants with their hands up. Although the authors were successful in creating a spiral wave, they noticed that the patterns did not follow computer simulations using the same rules. They noticed a breakdown of the spiral pattern that was attributed to variations in reaction time of the participants. Our cellular automaton is simpler than those proposed by Gerhardt *et. al.*[5-7] and Markus and Hess[8] and similar to the work of Greenberg and Hastings[9], extended to include random effects that mimic reaction times. The model of Greenberg and Hastings fell short in explaining various phenomenon seen in the literature. For example in the work of Huang *et al.*[10] spiral waves were seen drifting across the voltage sensitive dye imaging pads that they had set up. Rohlf *et al.*[11] explicitly included drifting in their work of spiral wave dynamics in excitable media in spheres. Qiang and Zhang recently developed a grid model to model spiral waves in an excitable medium[12]. In their work they explicitly included a diffusion term in addition to the inhibitor and activation terms. They were able to generate drifting spirals that are triggered by a single activated site. In our model diffusion is not explicitly included, but the randomization of the characteristic times induced a drifting of the initial spiral wave that behaved like a normal diffusive process. Our cellular automaton keeps the simplicity, elegance, and efficiency as it only allows local interactions. Nevertheless, there is a richness in the observed phenomena and we show that it is able to recreate naturally occurring effects such as the natural breakdown of the spiral wave patterns as time evolves. We observe similar phenomena as observed by Fenton *et. al.*[13] that explicitly studied different breakdown mechanisms in cardiac activity. We believe that allowing the system characteristic times in a model of interacting cells to fluctuate will lead to more realistic systems in which phenomenon seen in the literature naturally appear. We expect to convince the reader that the chosen simplified cellular automaton is capable of offering an in depth understanding of spiral wave formation and stability.

---

[1] Corresponding author: p-acioli@neiu.edu

## II. Method

### A. Model

Our cellular automaton implementation is general and it can represent any set of interacting units, such as neurons, species, or particles. Our focus in the present work is on a network of neurons where an electric signal is propagated. The pioneer work of Hodgkin and Huxley shows that the propagation of signals in the membrane of neurons can be modeled by an electric circuit that depends on a series of factors such as the ionic currents and membrane capacity[14]. In most cellular automaton models, the process of charging and discharging of the capacitor can be more simply represented by excitation and refractory times that imply that the neuron will get excited once a threshold voltage is exceeded and it will remain excited for the duration of the excitation time. After it gets de-excited it will remain in such state for the duration of the refractory period.

Our implementation takes advantage of this simplicity and is set up as a square grid of excitable nodes. Each node is in one of three states: ground, active, or refractory. Ground state nodes wait to be excited. Once a ground state node becomes excited it stays in this state for a period of time called the excitation time. Refractory nodes remain in this state for a refractory period when they decay into the ground state. Each node has an internal condition that once is met will trigger ground state nodes to activate. This internal condition, excitation threshold, is the number of neighbors that are in an excited state. The threshold value is set at the beginning of the simulation, and for the majority of the results presented here is set at two. Each node has a list of neighboring nodes that they interact. To improve efficiency the list of neighbors is determined at the beginning of the simulation by searching for nodes that fall within a given radius. If we consider the distance between nearest neighbors to be $a$, then a search radius of distance of $1.5a$ corresponds to the eight neighboring nodes seen in Fig. 1. We have tested different search radii and found that the value of $1.5a$ reproduced well the spiral waves reported in ref.[4]. We also explored adding weights to the nodes that were distance dependent. These lead to different width spiral waves with smoother patterns. In a traditional cellular automaton each excited node will remain excited for a fixed time, after which it will enter the refractory state for a fixed refractory period. In the present work we allow both the excitation and refractory times to fluctuate in a random fashion between pre-determined limits.

To validate our implementation, we first studied the conditions for formation and extinction of spiral waves using fixed excitation and refractory times, as well as varying the threshold condition for excitation and the effect of the search radius. Once these have been established, we studied the effects of allowing fluctuations of the characteristic times. One of the consequences of these fluctuations in the active and refractory periods is the break of a spiral wave and spawning of new ones. To help on our analysis of these effects we needed to track the center of the initial spiral wave as well as tracking the formation of new ones. In the next subsection, we describe how this is accomplished.

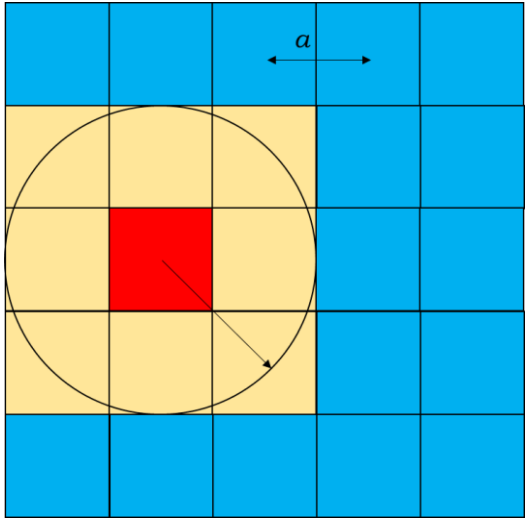

Figure 1: Representation of a node and its neighbors. The yellow squares represent the 8 neighbors of the highlighted node (red) using a search radius of 1.5*a*. Where *a* is the nearest neighbor distance.

B.  Defining a Spiral Wave

In this section, we describe our algorithm to identify spiral waves during the simulation. This is accomplished by searching the discrete grid and identify visually observed properties that are common to all spiral waves and quantifying them. During our initial simulations, we observed some commonalities in the spiral waves. First, a single spiral wave is composed of a continuous wave front except for its center. This observation provided us with a robust and simple definition of spiral wave. A continuous open wave that originates in a single point that propagates outward. That means, in practice, that we will search for breaks in the wave front of active states. The algorithm to search for these breaks and find the centers is described below.

To properly identify the spirals and their center, we need to define an edge cell in the wave front as an excited cell that has ground state cells as neighbors. The center of a spiral wave is an edge state that only has one edge state as its nearest neighbor, thus representing a break in the wave front. In Fig. 2 we show two examples of the search for the center of spirals. For every edge state, we search its neighbors, if there is only one neighbor cell that is an edge then that cell is a candidate for a center. We then extend the search to the second nearest neighbors to make sure that the wave front is continuous as indicated in Fig. 2a). In Fig. 2b) we show an edge state that does not represent a spiral wave center. In our code we used this algorithm to identify the center of the first spiral wave that is generated as well as to identify breaks in this initial wave that spawn additional spirals. We then tracked the number of breaks as a function of simulation time.

C.  Necessary conditions for spiral wave generation

Our preliminary search for initial conditions that allow for spiral wave formation leads to a common trend of a barrier of refractory states that are in contact with a front of excited states. This result is corroborated by literature on spiral waves[4]. The excited wave front will coil around the barrier and when the refractory barrier decays into ground state nodes the spiral wave pattern will form. If one creates a barrier of refractory states that does not decay, the wave front will close again in its original form and spirals are not formed. In Fig. 3 we have an example of a plane wave propagating from left to right and a long lived island of refractory states. On the second panel the wave front interacts with the island and starts coiling around. When the barrier decays it allows for the formation of a pair of spiral waves rotating in opposite directions.

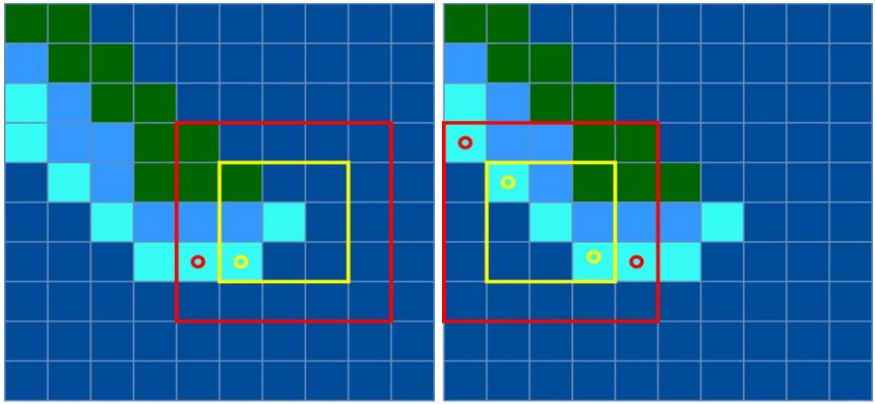

Fig. 2 a) Example of an edge cell that is identified as the center of a spiral wave. b) Example of an edge cell that is identified not to be a center of a spiral wave.

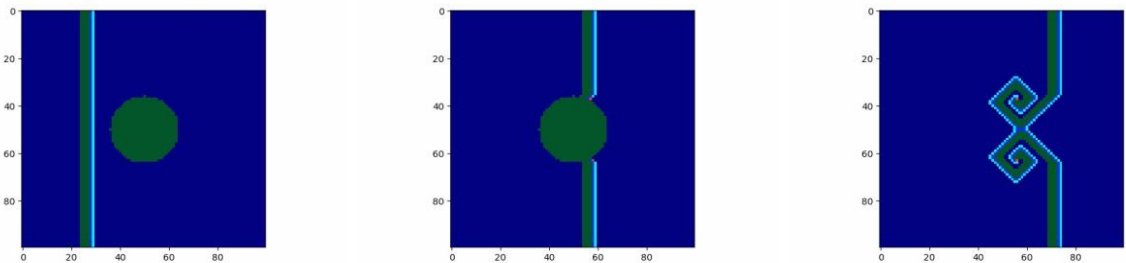

Fig. 3: A plane wave propagating through a medium with an island of long lived refractory states. In the first panel the wave is not yet interacting with the island. The second panel shows plane wave interacting with the island leading to a break on the wave front and the initial coiling. In the third panel the barrier decays and a pair of spiral waves is formed.

In order to study the factors that lead to formation of spiral waves and to determine the relationships between these factors and the characteristics of the spiral waves, we decided to use a standard initial condition of a line of excited states next to a line of refractory states, similar to the conditions reported in ref.[4]. As a benchmark to study the effects of the randomization of the refractory and excitation periods we run simulations with fixed refractory and excitation periods, given in units of simulation time steps. We noticed that the randomization of these times lead to several phenomena that are described below.

### III. Results

#### A. Stable Spiral Wave Generation

In order to test our code we tested it on conditions similar to those reported by Welsh *et al*[4]. We chose a 100x100 square simulation grid. We fixed the refractory time and the excitation (active) time at different values. We set a double column of 50 excited cells adjacent to a column of 50 refractory states as indicated in Fig. 4. This leads to a clockwise spiral pattern. In Fig. 5 we display the results of the spiral patterns for different values of the excitation ($t_{exc}$) and refractory ($t_{ref}$) times. In Fig. 5a) we set $t_{exc}=2$ and $t_{ref}=5$. One notices that the spiral wave of refractory states is wider than that of the excited states. In Fig. 5b) we set $t_{exc}=2$ and $t_{ref}=2$. In this case, the width of both wave fronts are the same. Finally, in Fig. 5a) we set $t_{exc}$=5 and $t_{ref}$=2, leading to the spiral wave of excited states being wider than that of the refractory ones. It is clear from these results that spiral waves can be generated with different combinations of characteristic times and that the width of the waves is directly proportional to these times, refractory or excited. We also observed that under these conditions, the spiral waves are stable and continue without degradation and the center of the wave stays stationary. As originally observed in ref. [4], one cannot explain the degradation observed in the human spiral waves under fixed conditions. Welsh *et al*[4] speculated that the degradation could be attributed to different reaction times by the participants and not accurately counting their active and refractory periods. To test this hypothesis we decided to implement in our algorithm a uniform fluctuation of the refractory and excitation times between pre-established bounds. In the next subsection we report the effect of changing the excitation threshold.

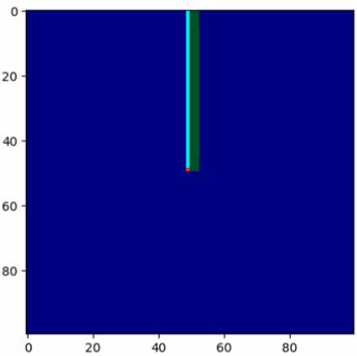

Fig. 4 – Initial condition used to generate a sable clockwise spiral wave. The green cells represent refractory states, the blue light blue represent excited states, the cyan represent excited edge states, and the dark blue represent the ground state cells

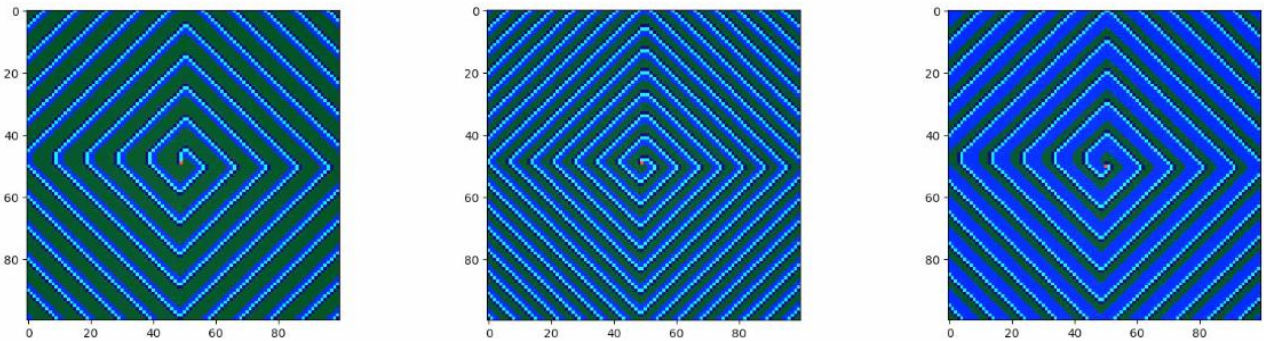

Fig. 5 – Spiral waves on a 100x100 grid with different excitation and refractory times. The green cells represent refractory states, the blue light blue represent excited states, the cyan represent excited edge states, and the dark blue represent the ground state cells. a) $t_{exc}$=2 and $t_{ref}$=5. B) $t_{exc}$=2 and $t_{ref}$=2. C) $t_{exc}$=5 and $t_{ref}$=2.

B. Effect of the Excitation Threshold

In the previous section we presented some of the conditions for the creation of stable spiral waves in a system analogue to a network of neurons. Here we present some results of how the trigger condition, threshold of excitation, affects the generation and shape of spiral waves. We fixed the excitation and refractory times at $t_{exc}$=5 and $t_{ref}$=2, and studied the effects of changing the excitation threshold ($N_{thr}$) to 1, 2 or 3 excited neighbors. The results are presented in Fig. 6. In each of the cases a clockwise spiral wave was formed. However, each had a different geometric shape. In the case that a single neighbor state was needed to trigger an excitation the spiral formed was square. When the trigger condition was two excited states, the spiral wave had a diamond shape, and for $N_{thr}$ = 3 and octagon was formed. If one requires that 4 neighbors are excited no waves will be propagated if one limits the search radius to only include the eight adjacent cells as neighbors (search radius = 1.5). If the search radius is increased to include more cells then $N_{thh}$ > 3 could lead to spiral waves, however this will result in a slower code without resulting in new phenomena. Therefore we limited our runs to $N_{thr}$ = 2 for the results that follow with a search radius of 1.5 unit cells.

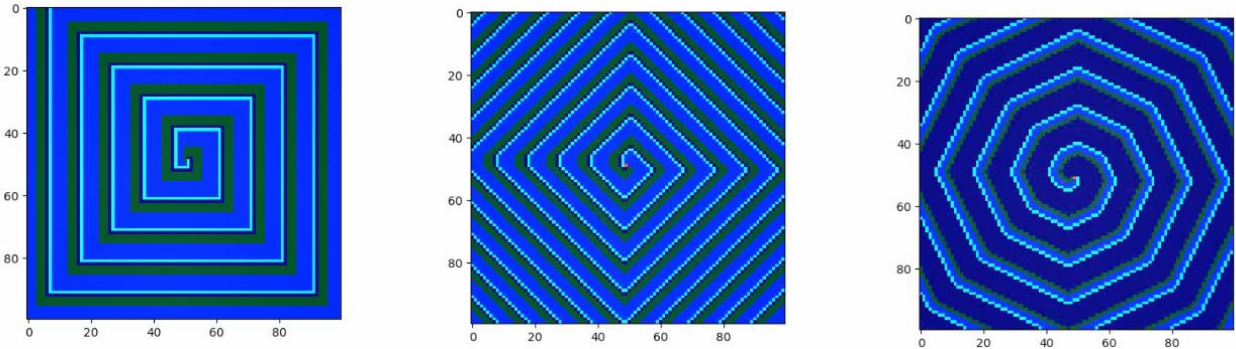

Fig. 6 – Spiral waves on a 100x100 grid with different threshold conditions. a) $N_{thr}$ = 1. b) $N_{thr}$ = 2. c) $N_{thr}$ = 3.

### C. Randomization of the Characteristic Times

As mentioned in the introduction our motivation was to accurately explain the observed breakdown of the human spiral wave of Welsh *et al.*[4] The algorithm described in the previous sections generates stationary stable spiral waves. The authors suggested that the breakdown was due to the human reaction times. We test this hypothesis by allowing the refractory and active times to fluctuate uniformly within a set range, similarly to the early work of Moe et al.[15] . These fluctuations not only accurately explain the breakdown to the human spiral wave, but, also lead to more interesting effects not observed in the case the characteristic times were fixed. Among these effects, we notice the drifting of the initial spiral wave, the breakdown of the spiral waves with concurrent spawning of multiple spirals, and a doppler effect. The breakdown of spiral waves have been observed in models of intracellular $Ca^{2+}$ dynamics[16], while drifting was observed in mice neural tissue[6]. In the present work we will explore these effects as being a consequence of the randomization of the refractory and excitation times.

When implementing the randomization of the refractory and active times we expected to observe the breakdown of the spiral waves, an unexpected effect we observed was the drifting of the center of the initial spiral wave. To quantify this drifting we tracked the position of the center of the initial spiral wave as a function of simulation time. We ran 100 different simulations, starting with the same initial conditions but with different random number seeds. We averaged the results to determine the average distance from the initial origin as a function of time. We also tracked the vertical and horizontal displacements. In Figs. 7a) and 7b) we plot the horizontal and vertical displacement of each individual run and in 7c) and 7d) we plot the average over all runs. One notices that in some cases an individual run is interrupted before the simulation is finished. This is a result of the system becoming too chaotic that we lost track of the center of the initial spiral wave. As we see from the average of all runs, the average displacement is about zero, with standard deviations that increase as a function of time. This is an indication that this is a normal diffusive process. In order to confirm this we plot the overall average displacement

$$\langle r \rangle = \left\langle \sqrt{x^2 + y^2} \right\rangle . \qquad (1)$$

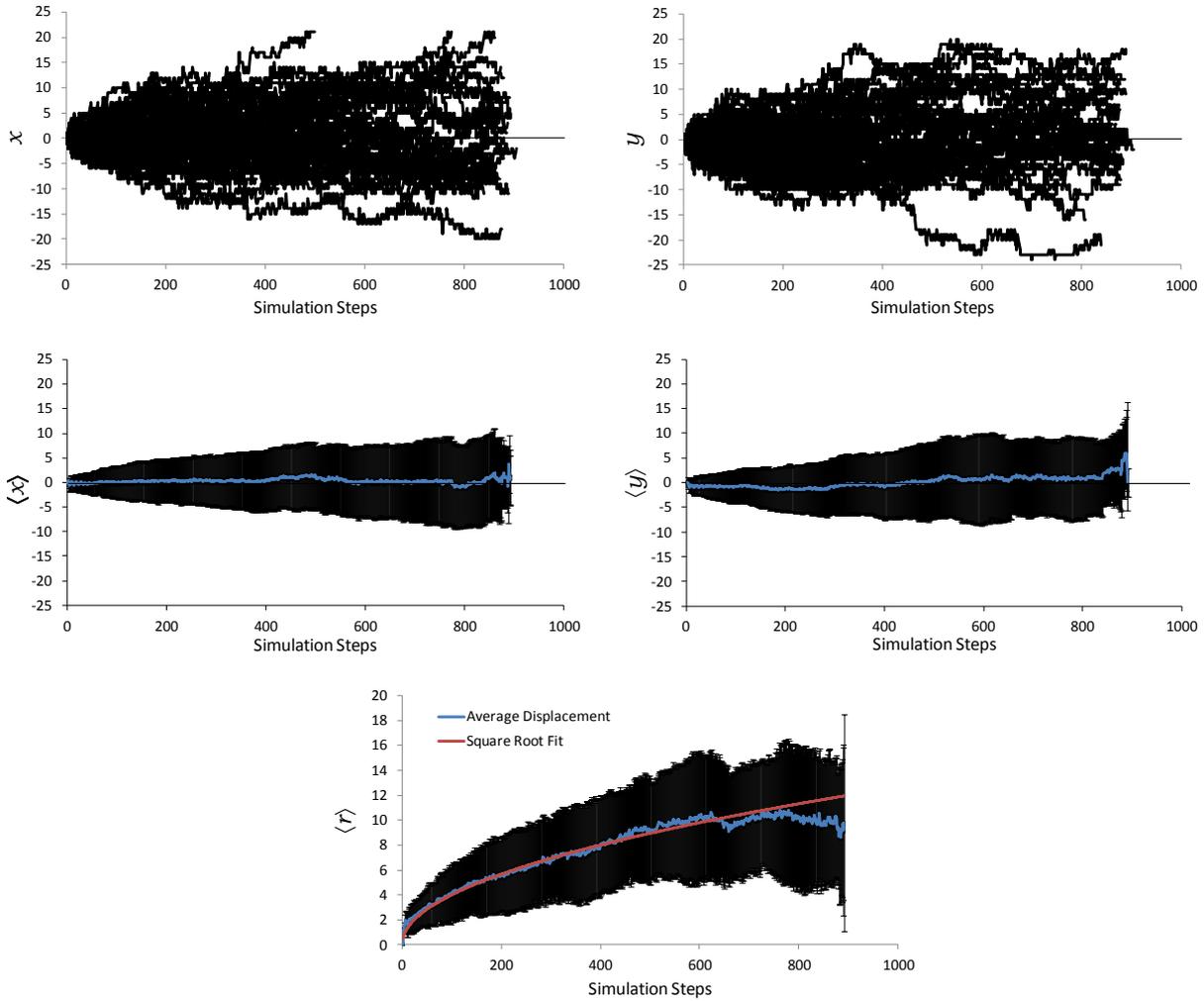

Fig. 7 – Center of spiral wave drift of 100 simulations with different random seeds and $t_{exc} = 1$-$2$, and $t_{exc} = 2$-$5$. a) Each line on the graph represents the horizontal displacement of an individual run. b) Each line on the graph represents the horizontal displacement of an individual run. c) Average horizontal displacement with standard deviation. d) Average vertical displacement with standard deviation. e) Average distance from original spiral wave center.

In a normal Brownian diffusive process the average displacement is proportional to the square root of the simulation time. In Fig. 7e) we plot this average and a square root fit that confirms the regular diffusive process. The deviations for large simulation times can be explained by the fact that in many cases we lost track of the center of the wave as a result of too many breaks on the original spiral wave. In conjunction with drift we observed the breakdown of the initial wave front spawning several spirals. in Fig. 8 we plot the spiral waves count for the same 100 runs presented in Fig. 7. As one can see as time evolves the effects of randomization increase evidenced by the increase of the number of spiral waves. We expect these effects to be more pronounced if we allow for a wider range of excitation and refractory times. To better understand this graph we present snapshots of one of the runs in Fig. 9.

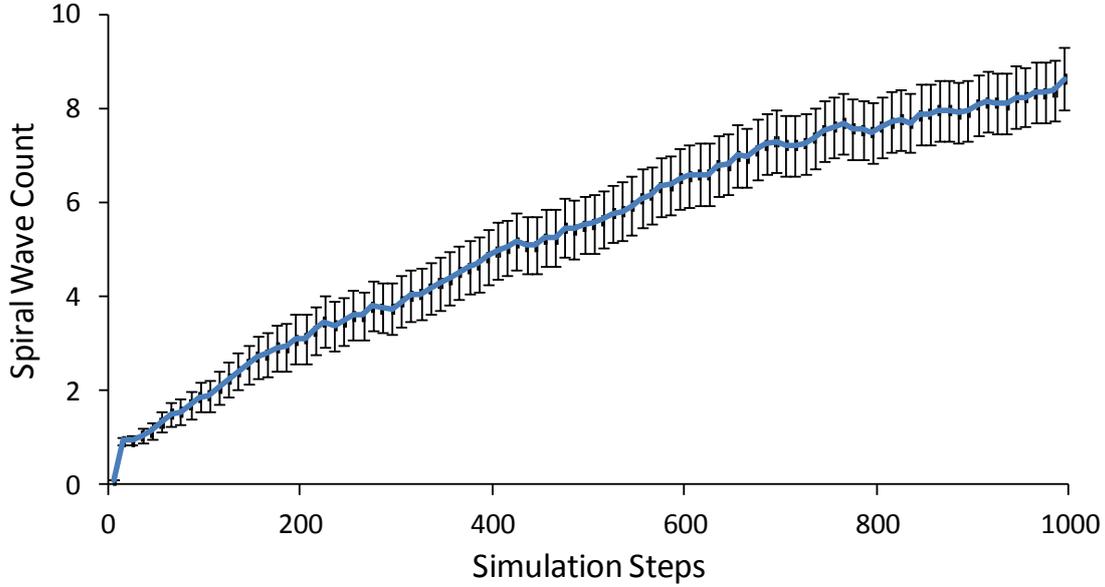

Fig. 8 – Average of spiral wave count of 100 simulations with different random seeds and $t_{exc}$ = 1-2, and $t_{refr}$ = 2-5.

In Fig. 9b) one can see that a single spiral wave is formed, but some of the effects of randomization are more clear as the refractory portion of the wave front shows some small breaks. As time evolves one starts to detect breaks on the wave front spawning new small spirals. The red dots in Fig. 9c) are the centers of these spirals. In Fig. 9d) more spirals are formed and it is clear that the center of the initial spiral has drifted to the upper right corner of the simulation grid. Accompanying the drift of the spiral wave is a Doppler effect as the wave fronts in the direction of the drift tend to bunch together. This Doppler effect was also observed in the work of Fenton *et. al.*[13]

It is clear that Figs. 7-9 show that randomization of the characteristic times leads to a breakdown of the spiral waves. We now turn the discussion to the correlation of the range of randomization and the breakdown of these waves. This is accomplished by varying these ranges and quantifying the number of breaks and how fast they take place. We have noticed that spiral waves are more stable when the refractory period is longer than the excitation period.

In Table I we present the ranges of refractory and excitation periods we considered and the average number of breaks in the wave front once the system reaches equilibrium. Once again, these results are averaged over 100 runs with different random seeds. One can notice the spiral waves are more stable when there is no overlap between the ranges of active and refractory times. In some cases the system did not reach equilibrium in 1000 simulation steps. To better understand the data in Table I we plot the number of breaks in the wave front. In Fig. 10 we plot the cases where there is a large number of breaks to the point that one cannot recognize a spiral wave ($1 \leq t_{exc} \leq 2$, $1 \leq t_{refr} \leq 2$ ; $1 \leq t_{exc} \leq 3, 1 \leq t_{refr} \leq 3$; $1 \leq t_{exc} \leq 3, 2 \leq t_{refr} \leq 4$). These are the cases where there is a large overlap between the excitation and refractory periods. It is very clear that there is a quick breakdown of the initial spiral wave and clearly reaching an equilibrium state.

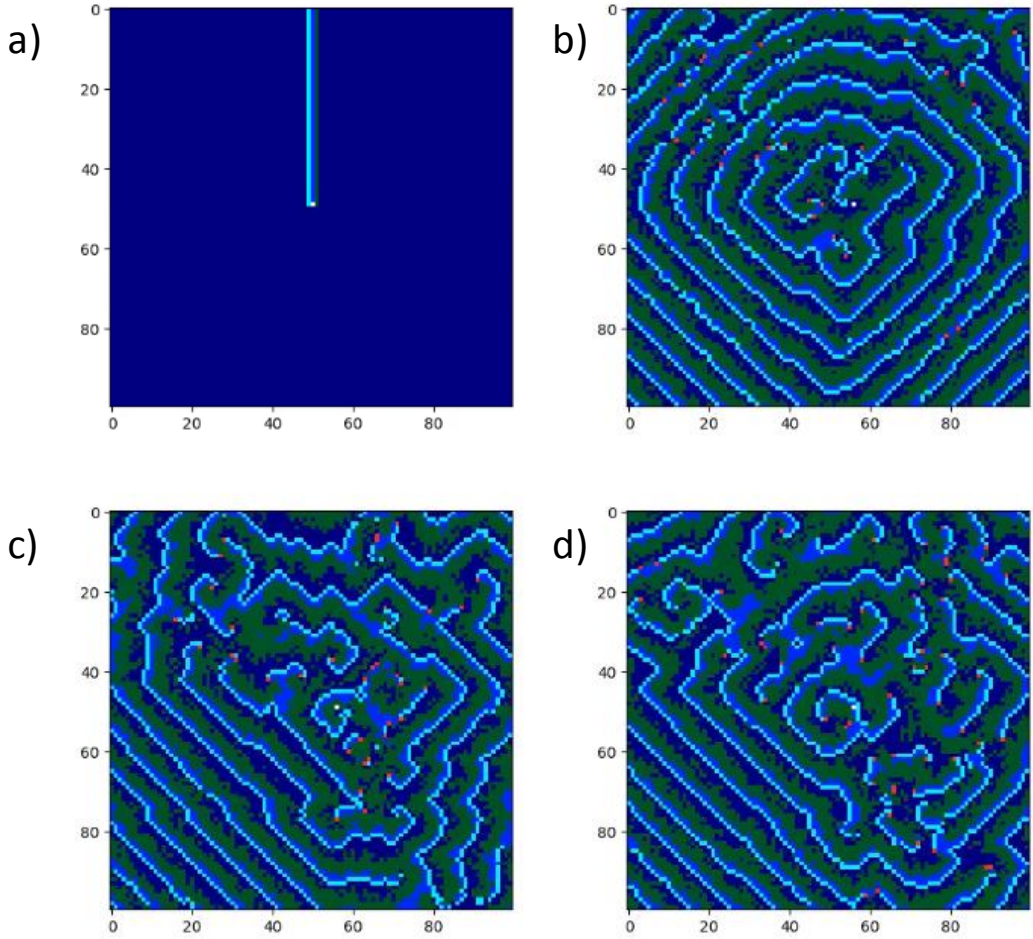

Fig. 9 – Snapshots of one of the simulations with $t_{exc}$ = 1-2, and $t_{ref}$ = 2-5 on a 100x100 simulation grid. a) Initial condition. b) At 333 simulation steps. c) At 667 simulation steps. d) At 1000 simulation steps.

Table I – Number of breaks in the wave front as a function of the excitation and refractory period ranges. $t_{exc}$ (min) and $t_{refr}$ (min), are the minimum value of the excitation and refractory periods, respectively. $t_{exc}$ (max) and $t_{refr}$ (max), are the maximum value of the excitation and refractory periods, respectively.

| $t_{exc}$ (min) | $t_{exc}$ (max) | $t_{refr}$ (min) | $t_{refr}$ (max) | Number of breaks | Obs |
|---|---|---|---|---|---|
| 1 | 2 | 1 | 2 | 151.8 | |
| 1 | 2 | 2 | 3 | 17.4 | Did not reach equilibrium |
| 1 | 2 | 3 | 4 | 2.7 | |
| 1 | 3 | 1 | 3 | 136.5 | |
| 1 | 3 | 2 | 4 | 150.9 | |
| 1 | 3 | 3 | 5 | 37.7 | Did not reach equilibrium |
| 1 | 3 | 4 | 6 | 7.5 | |
| 1 | 3 | 5 | 7 | 5.1 | |

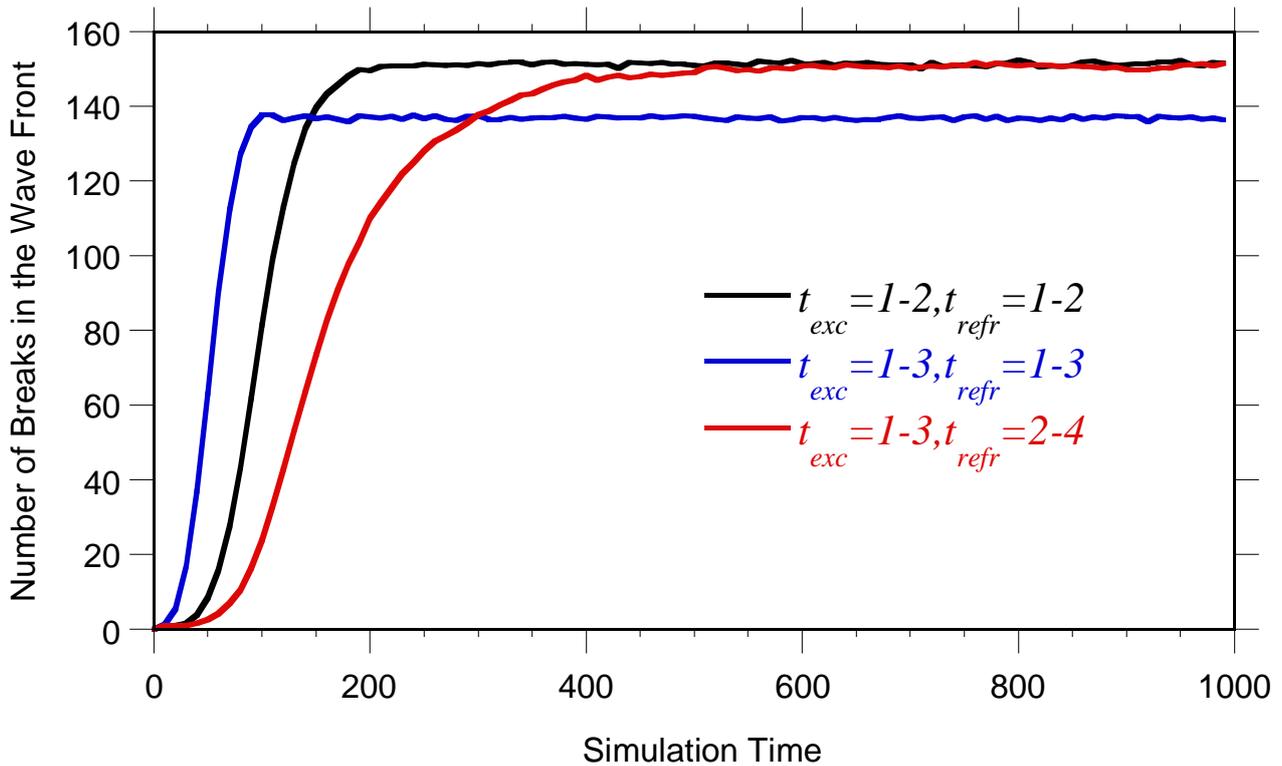

Fig. 10 – Number of breaks in the wave front for different ranges of excitation and refractory periods.

In Fig. 11 we plot the results for a range of active and refractory periods with an intermediate overlap. It is interesting to note that in both cases it seems like the number of breaks in the wave front continue to increase, but after 1000 steps the system has not equilibrated yet. A different picture is displayed in Fig. 12. In the ranges of the excitation and refractory periods with no overlap, the spiral waves are very stable, only exhibiting a small number of breaks in the wave front. It seems that the longer the refractory periods are the more stable the spirals are. To better visualize the results in Figs. 10-12 we plot in Fig. 13 the snapshot of each simulation after 1000 simulation steps. It is quite clear that in the cases where there is a complete overlap between the refractory and active periods the system becomes very chaotic as indicated in Figs. 13a) and 13c). As the overlap decreases, the system becomes more stable. Some of the more interesting cases occur at intermediate overlaps, from Figs. 13b) and 13f) one can see that at 1000 steps the system preserves some of the characteristics of the initial spiral wave, but it continues to breakdown at a slow pace. It is reasonable to expect that they might evolve to a case where there is no resemblance of a spiral wave, but one can still identify some wave fronts, such as in the case of Fig. 13e). When there is no overlap at all, the spiral waves are indeed very stable, showing an occasional break in the wave front, but not spawning a new spiral wave pattern. To test our hypothesis that the intermediate overlap might lead to a more chaotic behavior we runs the cases $1 \leq t_{exc} \leq 2$, $2 \leq t_{refr} \leq 3$ ; $1 \leq t_{exc} \leq 3$, $3 \leq t_{refr} \leq 5$, for a total of 10000 simulation steps.

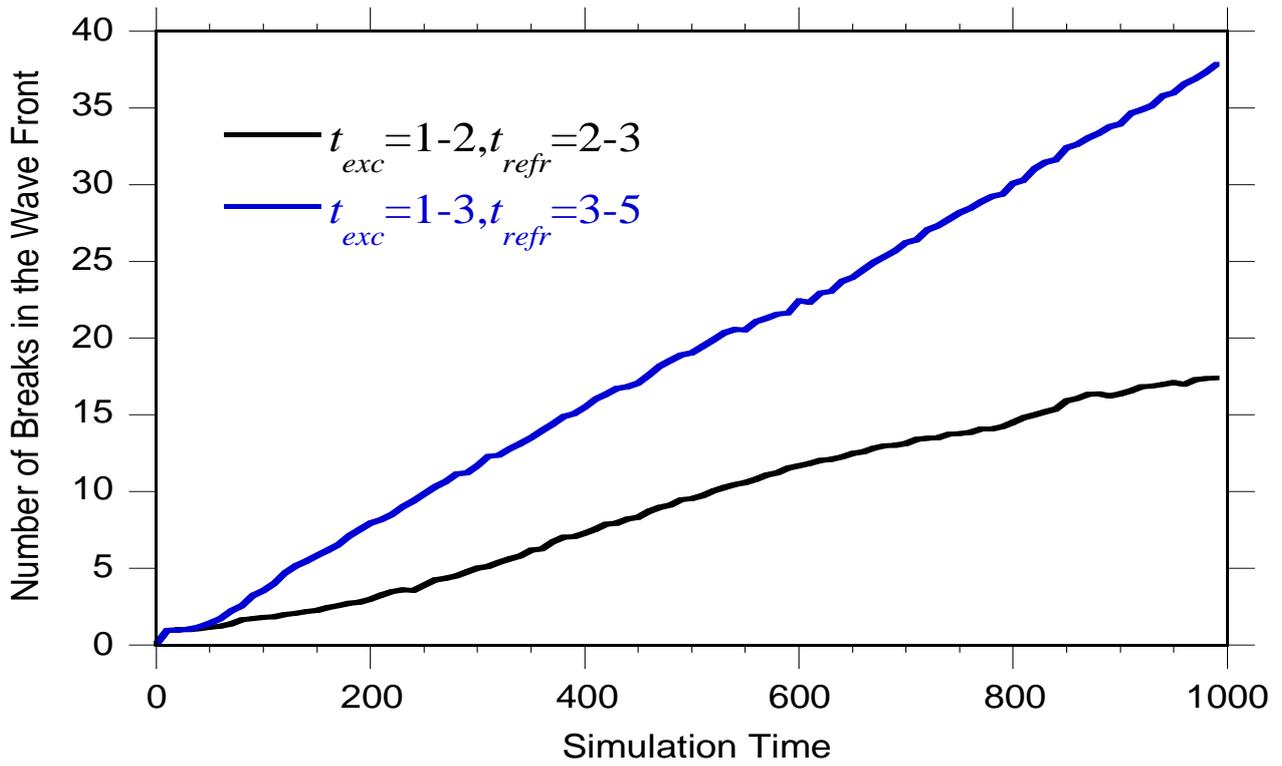

Fig. 11 – Number of breaks in the wave front for different ranges of excitation and refractory periods.

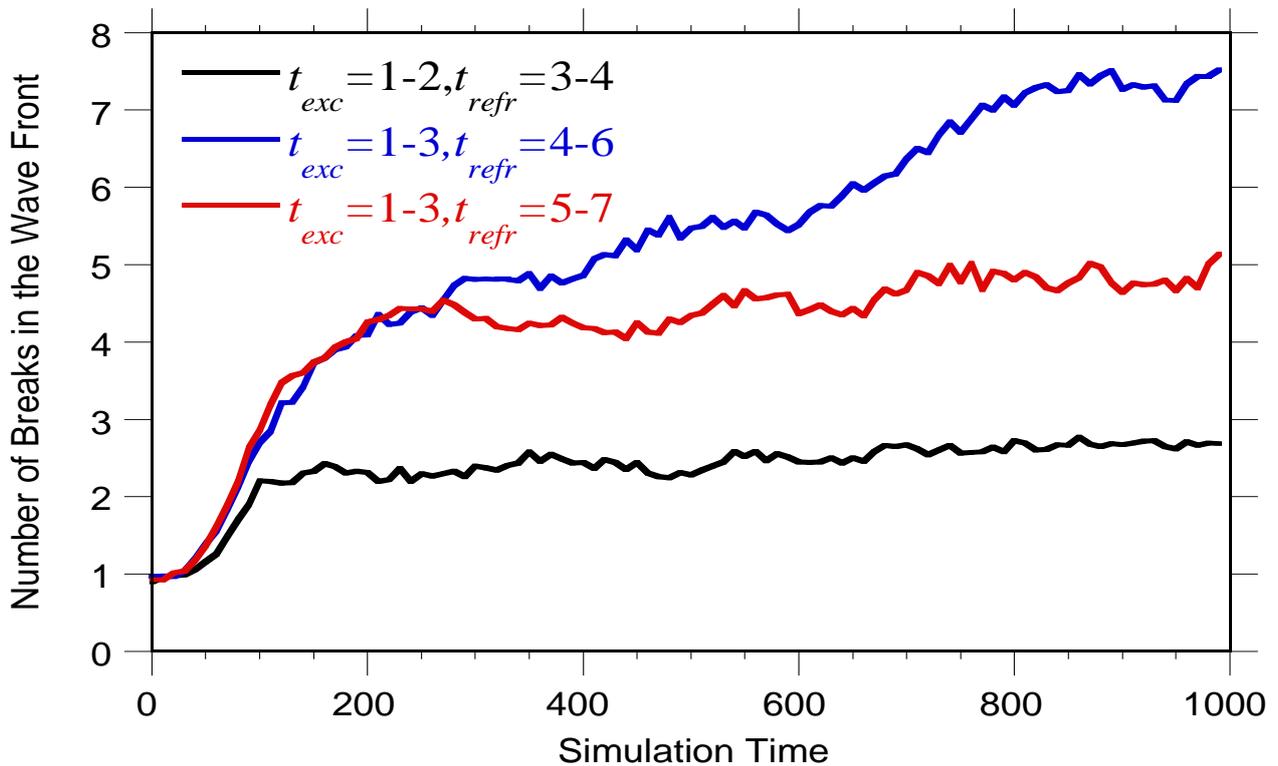

Fig. 12 – Number of breaks in the wave front for different ranges of excitation and refractory periods.

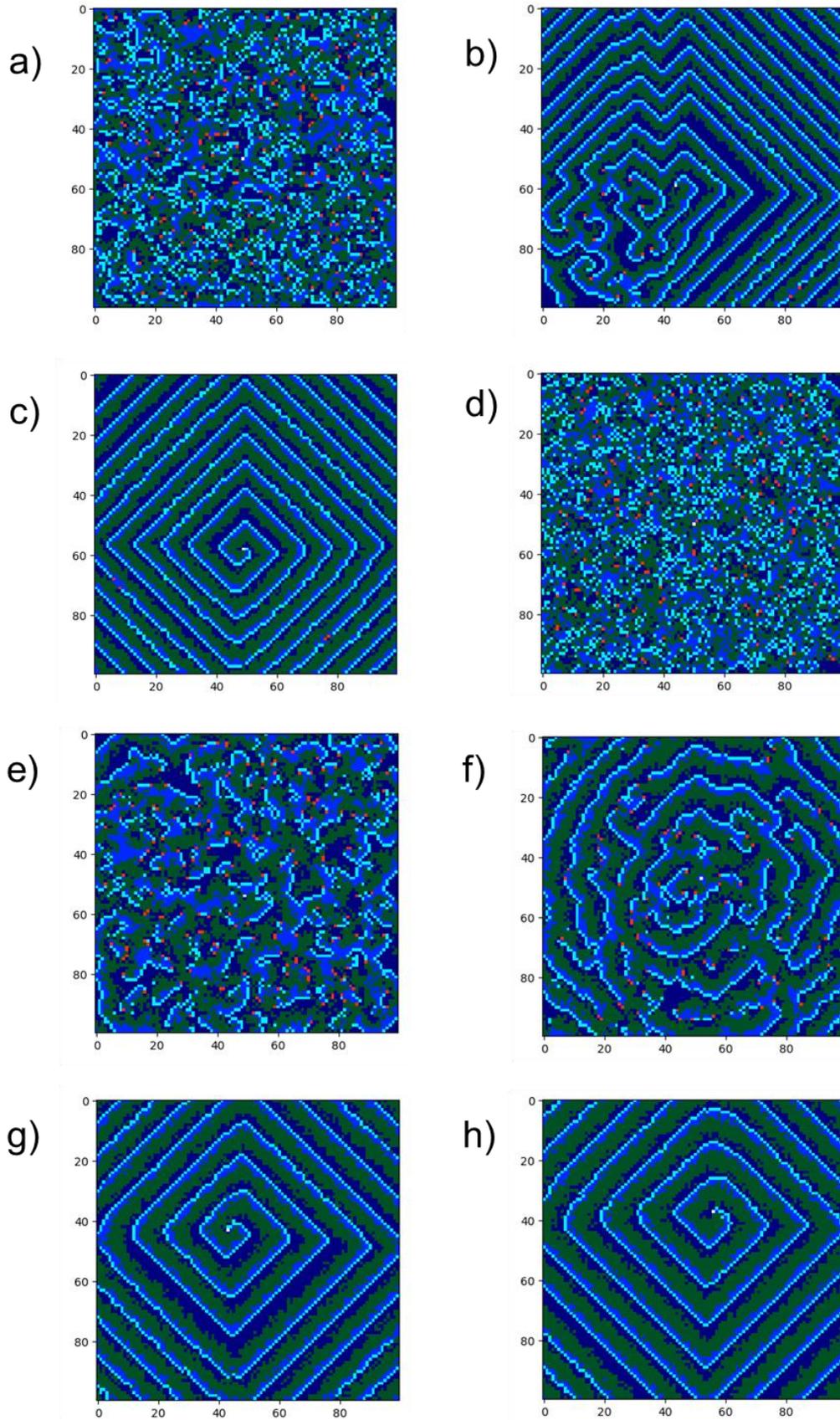

Fig. 13 – Snapshot of the 1000th step of each simulation presented in Table I. a) $t_{exc}$=1-2, $t_{refr}$=1-2. b) $t_{exc}$=1-2, $t_{refr}$=2-3. c) $t_{exc}$=1-2, $t_{refr}$=3-4. d) $t_{exc}$=1-3, $t_{refr}$=1-3. e) $t_{exc}$=1-3, $t_{refr}$=2-4. f) $t_{exc}$=1-3, $t_{refr}$=3-5. g) $t_{exc}$=1-3, $t_{refr}$=4-6. h) $t_{exc}$=1-3, $t_{refr}$=5-7.

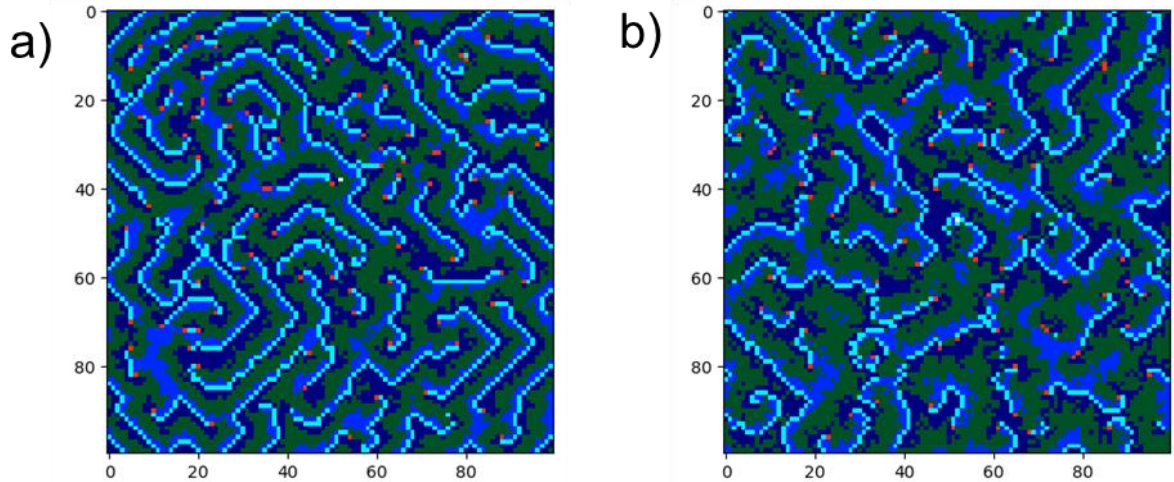

Fig. 14 – Snapshot of the 10000$^{th}$ step of the simulations for the cases of intermediate overlap. a) $t_{exc}$=1-2, $t_{refr}$=2-3. b) $t_{exc}$=1-3, $t_{refr}$=3-5.

One can see that the two cases of intermediate overlap have similar features as they seem to have lost most of the features of the initial spiral wave. However, this equilibrium seem to feature a number of smaller spirals, unlike the cases of strong overlap cases where the equilibrium seem to have no distinct features but a random distribution of active and refractory states. Interestingly enough, in none of the cases considered in this study there is a complete suppression of the active states.

## IV. Conclusion

We implemented a simple cellular automaton to propagate signals in an excitable medium. Similar to other models in the literature we describe the excitations by simple interactions between nearest neighbors that have characteristic active (excitation) and refractory periods. Because we define the list of neighbors at the beginning of the simulation our method is very fast as we do not have to search the simulation cell for its neighbors when deciding if a cell in the medium will be excited or decay to a refractory state. Under fixed excitation and refractory periods, we noticed that in order to produce spiral waves an active wave front must meet a barrier of refractory states that will allow the front to coil and start the spiral pattern. Once a spiral wave is started, it will remain stationary and stable for a long period of time.

We also study the effects of the threshold condition and of the search radius in the shape of the spirals. We noticed that for a case where the search radius yields 8 adjacent neighbors we were unable to generate spirals waves if the threshold condition was larger than 3 neighboring excited states. Since no different phenomena was observed we limited our simulations to $N_{th} = 2$. This allowed to fast simulations with on the fly animations for quick analysis.

The main focus of the present work was on the effects of randomization in explaining the results of the breakdown of a human spiral wave of Welsh et al.[4] The randomization was introduced in the active and refractory periods that were sampled from uniform distributions in a pre-determined range. When the ranges of active and refractory periods are identical, the system quickly evolve to an equilibrium state with no traces of spiral waves in what seems like a chaotic behavior. For an intermediate overlap of the active and refractory periods, the system evolves slowly to an equilibrium where one cannot identify the initial spiral wave, but one can observe several small spiral waves. If there is no overlap between the active and refractory periods, then the initial spiral wave is very stable and one only observed some breaks in the wave front but no secondary spiral waves are observed. Interestingly enough was that, although no diffusive terms were added to the model, a drift of the initial spiral wave was induced. In the cases where a spiral wave pattern was observed we were able to show that the drift followed a normal diffusive pattern where the average distance from the initial position was proportional to the square root of simulation time.

This simple implementation of randomness is able to properly reproduce the results of the human spiral wave and in the range of values of the excitation and refractory periods it might not explain the breakdown in biological systems given that in some cases the range are larger than the 25 percent of the set value. We believe that this can be remedied by using a range of refractory times like $t_{refr}$=9-10, reducing the difference to more reasonable values and we will still be able to observe the same kind of phenomena. We also plan to generalize the model to an excitation condition that is distance dependent to see if new phenomena are observed. Although sampling the ranges of excitation and refractory periods from a uniform distribution was able to reproduce the results of ref. [4] we plan to sample it from a normal distribution with different variances and determine the effects on the observed patterns. In addition, we plan to extend the model to simulate population dynamics and observe if there is any correlation between patterns and persistence of the populations.

Acknowledgments

We would like to thank funding from Northeastern Illinois University Student Center for Science Engagement (SCSE) and the U.S. Department of Education (USDOE) Title III Award # P031C160209.